# From Double Colloidal Networks to Core–Shell and Mixed Composites through Sequential Gelation

*Alexander Kaltashov[1], and Safa Jamali[1,2*]*

Dept. of Chemical Engineering, Northeastern University, Boston, MA 02115
Dept. of Mechanical & Industrial Engineering, Northeastern University, Boston, MA 02115
*Email address: s.jamali@northeastern.edu

Multicomponent gel systems have garnered much interest due to their compelling mechanical properties in the past decade. Yet, some mechanisms associated with multicomponent gels, such as sequential gelation, have been explored primarily in the context of chemical non-reversible polymeric and protein gels than in physical reversible colloidal ones. In this study, we use mesoscale simulation techniques to model the sequential gelation of two-component colloidal systems whose components' interspecies and intraspecies electrostatic interactions can be modified independently. We show that by simply leveraging temporal control and interspecies interactions, we can construct markedly different networks; from double networks to mixed and core-shell composite structures of varying coarseness and heterogeneity natures. These findings present a compelling case for further exploration of multicomponent colloidal systems.

## 1 INTRODUCTION

Gels represent a relatively large class of soft materials with applications ranging from food(1) and cosmetics(2) to pharmaceutical products(3). These semi-solid materials are characterized by networks of interconnected components, constructed from a variety of building blocks such as polymers, proteins and colloidal particles(4). In the case of colloids, physical gels form as individual particles form physical bonds and progressively assemble into space-spanning structures (5–13). Although the overwhelming majority of microstructural and dynamical studies on colloidal gels to date are focused on single component systems, many real-world systems of interest involve more than one type of constituent. Compared to single-component systems, gels with multiple constituents allow for great structural diversity, giving rise to what are often appealing, non-intuitive mechanical properties (14–16). For instance in chemical gels, designing polymer network with distinct length scales can lead to formation of sacrificial networks with superior mechanical properties compared to gels made by any of the individual polymers(17–20). Nonetheless, similar pathways to designer double networks in physical gels are generally lacking.

The structural diversity and compelling properties observed in multicomponent systems is due in no small part to the introduction of new, tunable parameters that are unavailable in single-component systems. It is known, for example, that the nature of the two-body interactions between different constituents play a key role in the structure of multi-component systems by determining the extent of constituent mixing or separation into distinct networks (21, 22). An important mechanism that elevates the level of complexity in multicomponent systems compared to single component gels, is sequential gelation. Sequential gelation occurs in multiple stages rather than one, often through the exploitation of independent gelation mechanisms. Typically, this involves changes in environmental conditions such as pH, temperature or ionic strength(23–27), which trigger the gelation of components in a sequence. Although to date, some work has been done on multicomponent colloidal gels undergoing sequential gelation(24–26), in practice such systems are still relatively underexplored in comparison to the vast body of research concerning multicomponent polymer and chemical hydrogels undergoing such formation processes(28–32). Yet, the variety of directed self-assembly mechanisms available to colloids(33, 34) opens up compelling prospects for the construction of multicomponent colloidal gels via sequential gelation processes. Exploring temporal control in the context of attractive colloidal systems can potentially enable structures and properties that are not simply accessible through conventional gelation mechanisms. In this study, we examine the interplay between the strength of colloidal attractions, and a secondary gelation delay in simulated two-component colloidal gels undergoing sequential gelation. Hence, in the studied systems, one type of colloid undergoes gelation while the other component remains Brownian within the mixture during that time. Then, and following a pre-determined gelation delay, the local environment is changed in such a way that allows the second colloid species to associate the initial structure and undergo gelation as well. Figure 1A-C shows the schematic view of this two-step gelation process (**Fig 1A-C**). With a goal of better understanding the resulting morphologies and also the kinetics of gelation process, we evaluate the lasting impact of secondary gelation delay on the resulting structures by characterizing both the evolution of the particulate network and the final morphologies of the studied colloidal systems at long times. In particular, we illustrate how the impact of the gelation delay depends largely on the nature of the interspecies interactions present in the system.

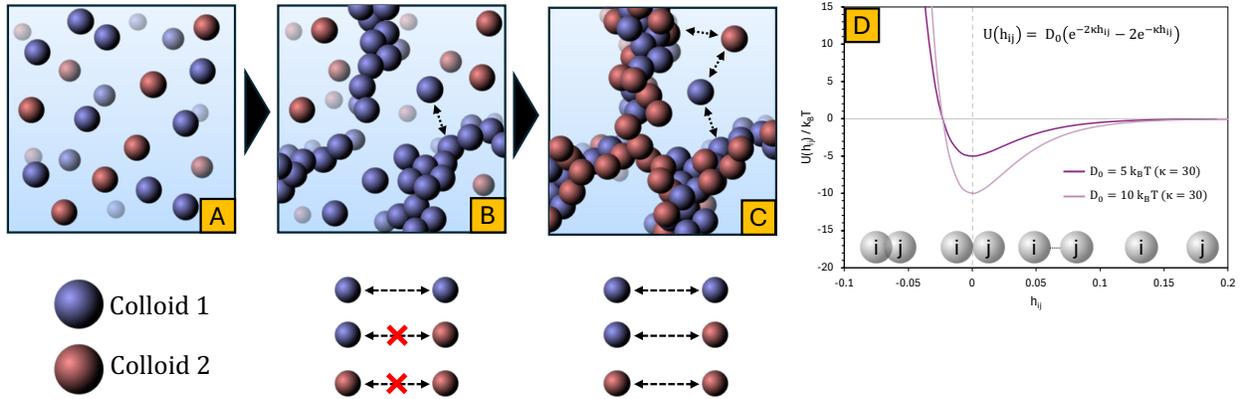

**Figure 1**: (A-C) Visual representation of the studied sequential gelation process (A) Colloid 1 ($C_1$) and Colloid 2 ($C_2$) populations are randomly dispersed in solvent. (B) $C_1$ particles begin experiencing intraspecies attraction and form an initial structure while $C_2$ particles remain non-interactive. (C) After a delay, $C_2$ particles begin experiencing interspecies and intraspecies attractions and begin interacting with the initial structure. (D) The short-ranged Morse potentials used to model particle interactions. Note that non-interacting particles are modelled to experience only the "repulsive" term of the Morse potential when overlapping, and a potential of $U(h_{ij}) = 0$ otherwise.

## 2 RESULTS

In this work, we developed and implemented a two-stage simulation protocol. First, the early stages of self-assembly were modelled using Dissipative Particle Dynamics. Next, the simulation box was scaled across its periodic boundaries, and the subsequent evolution was modelled using Langevin Dynamics. The durations of the two stages were fixed and identical in all systems studied, and were independent of the time at which the second colloidal species began interacting with the initial scaffold; this interaction could commence in either of the two stages. This simulation methodology, along with the rationale for its use, is described in detail in the Materials and Methods section.

The population of colloidal particles modelled in this study remained consistent throughout both stages of the simulation. The particles in question are evenly split into two species: Colloid 1 ($C_1$) and Colloid 2 ($C_2$). Aside from belonging to different species, these two particle populations are otherwise identical; with both possessing the same density and mean hydrodynamic radius of $a = 1.0 \pm 0.05$. Polydispersity was introduced into both populations' particle radii to suppress the crystallization that readily occurs in models neglecting hydrodynamic interactions (8, 35–37). We use short-range Morse potential to model short range attraction and nearly hard-sphere repulsion (**Fig 1D**):

$$U(h_{ij}) = D_0\left(e^{-2\kappa h_{ij}} - 2e^{-\kappa h_{ij}}\right)$$

Where $D_0$ represents the attraction strength, $\kappa^{-1}$ represents attraction range (in all cases, we used $\kappa = 30$) and $h_{ij}$ describes the interparticle surface-to-surface distance. Three distinct Morse potentials are considered in this study; the intraspecies attraction between $C_1$ particles ($D_{0\,(1-1)}$), the intraspecies attraction between $C_2$ particles ($D_{0\,(2-2)}$), and the interspecies attraction between $C_1$ and $C_2$ particles ($D_{0\,(1-2)}$). All simulations are described in terms of $a$, $k_BT$ and $\tau_D = (6\pi\eta a^3)(k_BT)^{-1}$, the dimensionless characteristic scales for distance, energy and diffusion time of a colloid, respectively. Here, $k_B$ is the Boltzmann Constant, $T$ represents temperature, and $\eta$ represents fluid viscosity. In studied systems, $C_1$ particles began experiencing intraspecies attraction at the start of every simulation. In contrast, $C_2$ began as non-interactive particles, and only begin experiencing intraspecies and interspecies attractions after a "secondary gelation delay" ($t/\tau_{D\ delay}$) (**Fig 1A-C**).

In the studied systems, $t/\tau_{D\ delay}$ was be set to either 20, 100, 200 or 3,000. In all studied systems, both intraspecies attractions were modelled to be identical and relatively weak ($D_{0\ (1-1)} = D_{0\ (2-2)} = 5k_BT$). However, interspecies attractions could be either non-existent ($D_{0\ (1-2)} = 0$), weak ($D_{0\ (1-2)} = 5k_BT$), or strong ($D_{0\ (1-2)} = 10k_BT$). Simulations were performed for every combination of $t/\tau_{D\ delay}$ and $D_{0\ (1-2)}$. In our characterization, we focused on determining how modifying these parameters would affect both the overall gel structure and mixing of the two different species.

The final gelation states at 5,500 $\tau_D$ are shown in **Fig 2** for all systems studied. To facilitate visual inspection of the resulting structures, each snapshot is presented mono-colored as well as bi-colored ($C_1$ in blue and $C_2$ in red): while the colloidal domain sizes are more easily distinguishable in mono-colored images, the level of mixing/de-mixing is more distinct when shown in different colors. By adjusting the $D_{0\ (1-2)}$ and $t/\tau_{D\ delay}$, a variety of morphologies are clearly formed; from double networks to mixed and coated structures. Visual observation reveals that changes in $t/\tau_{D\ delay}$ have a pronounced effect on the coarseness of the resulting microstructures in gels with a strong interspecies attraction ($D_{0\ (1-2)} = 10k_BT$). In contrast, changes in $t/\tau_{D\ delay}$ result in insignificant variation in the coarseness of systems with weak ($D_{0\ (1-2)} = 5k_BT$) or non-existent ($D_{0\ (1-2)} = 0$) interspecies attractions. This is somewhat expected: for the non-attractive systems of $D_{0\ (1-2)} = 0$, the system is expected to de-mix regardless of the gelation delay. As evident from the snapshots in Fig.3 (top row), all studied systems form double networks that are fully disconnected from one another. By introducing the same attraction as the one between each set of colloids ($D_{0\ (1-2)} = 5k_BT$) the system begins to mix with little to no change in the final domain sizes with varying values of gelation delay. Nonetheless, the blue versus red domains very clearly change as the gelation delay is prolonged, resulting in a core-shell composite structures at the longest delays studied. In order to quantitatively study the overall structures as well as the connectivity between different components, we have used a series of cluster and global level characterization techniques.

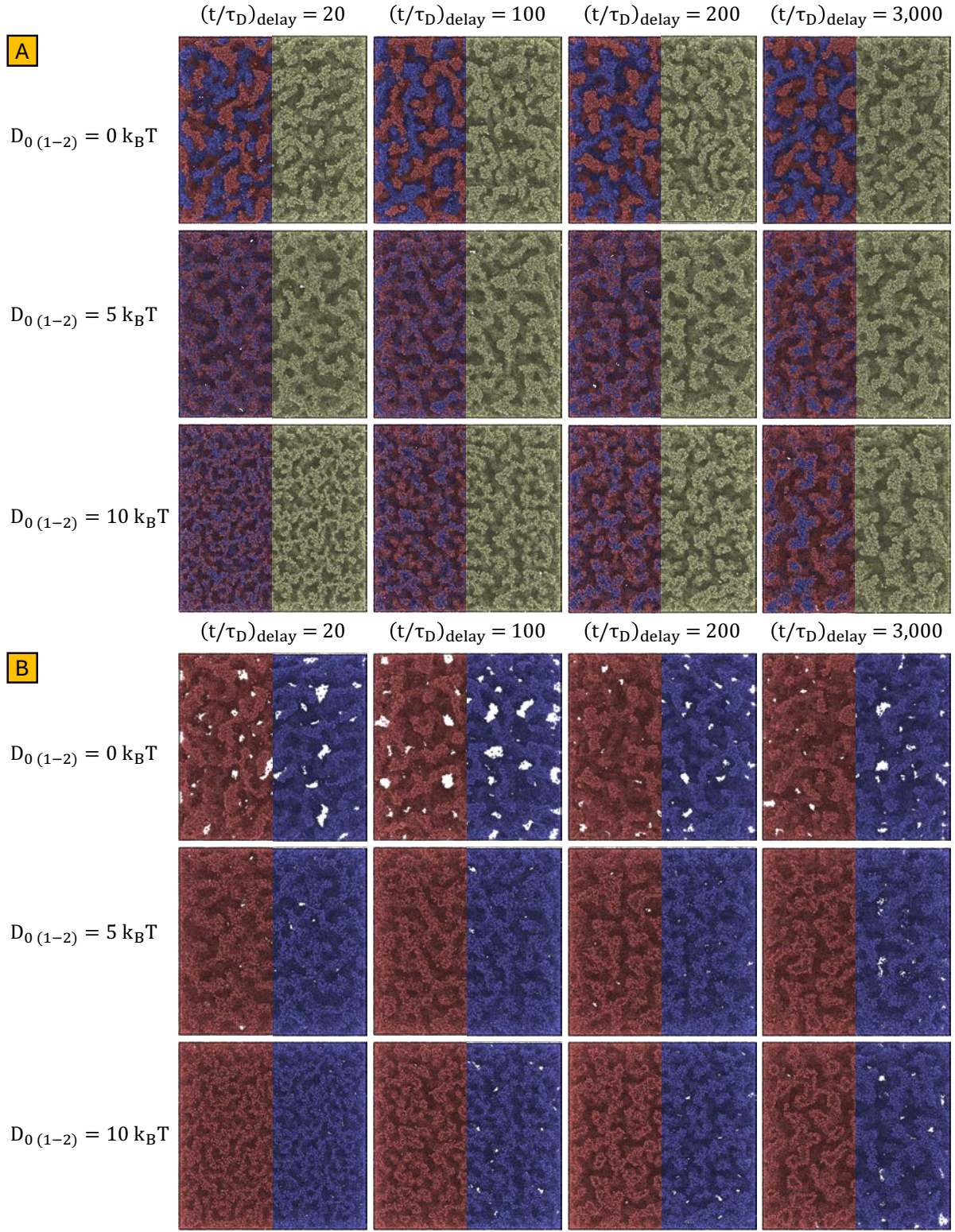

**Figure 2**: (A) Snapshots of the final structure of the simulated gels. (Left halves of snapshots) $C_1$ and $C_2$ shown in blue and red, respectively. (Right halves of snapshots) Both colloid species are colored green to illustrate overall gel structure. (B) Snapshots of the final structure of the simulated gels showing only (Left halves of snapshots, red) $C_2$ particles and (Right halves of snapshots, blue) $C_1$ particles.

## 2.1 Multi-Scale Structural Characterization

### 2.1.1 Static Structure Factor

We examine static structure factor $S(q)$, to gain insight into structures formed at the cluster level. Structure factor as a function of wavenumber $q$ can be approximated as the following(38):

$$S(q) = 1 + 4\pi\rho \int_0^\infty r^2(g(r) - 1)\frac{\sin(qr)}{qr}dr$$

Here, $g(r)$ is the radial distribution function (calculated using the Freud python package(39)), and $\rho$ is the colloidal particle density. A large peak in the range of $0.1 < qa < 0.4$ is present in every studied gel, indicating an abundance of structures of length scale $16 < l/a < 63$. The characteristic feature length $L_{S(q)} = 2\pi/q_{max}a$, gleaned from the peak position, is an indicator of the average size of clusters, i.e. the dominant length-scale of the overall structure. Aside from the peak's position, the intensity of the peak is also a measure of the number of particles in the clusters(40).

In gels with $D_{0\,(1-2)} = 10k_BT$, we observe two concurrent trends with increasing $t/\tau_{D\,delay}$: the shift of the peak to lower $q_{max}a$ values, as well as an increase in peak intensity. These findings clearly show that increasing the duration of secondary gelation delay results in the formation of larger, more dominant clusters in these systems. In contrast, systems with $D_{0\,(1-2)} = 5k_BT$ do not display significant variations in either peak intensity or in $q_{max}$ positions, indicating a lack of correlation between features sizes and $t/\tau_{D\,delay}$ (**Fig S1**). Systems with $D_{0\,(1-2)} = 0$ likewise do not display significant variations in peak intensity. Despite this, there is a decrease in the characteristic feature length at high $t/\tau_{D\,delay}$ values (**Fig S2**). Though it may be tempting to consider this as evidence of variation between the studied $D_{0\,(1-2)} = 0$ systems, other structural analyses (discussed later) fail to discern significant differences between their final states.

### 2.1.2 Pore Size Distribution

In addition to analyzing domains at the cluster level, we measure the distribution of interstitial pore size in the final structure, using the methodology proposed by Gelb and Gubbins(41). This measure not only provides the overall length scale of the pores within the particulate network structure, but it also yields their distribution as a measure of structural heterogeneity within a system. Here, a probe point is randomly chosen, and the diameter of the largest possible sphere encapsulating this point without overlapping with any particles is measured. Measurements were taken for a total of 50,000 probe points. It was found that in gels with $D_{0\,(1-2)} = 10k_BT$, increasing $t/\tau_{D\,delay}$ results in a pronounced increase in both the average pore size and the pore size distribution (**Fig 3A**). In contrast, minimal variations in porosity were seen in gels with $D_{0\,(1-2)} = 5k_BT$ or $D_{0\,(1-2)} = 0$ (**Fig S1, S2**).

### 2.1.3 Solvent-Accessible Surface Area

To assess the structure coarseness globally, Solvent-Accessible Surface Area (SASA) was measured for each system using analytical tools built into the VMD molecular graphics program(42). SASA is a measure of the surface area swept out by the center of a probe sphere. Here, the probe was selected to have a radius of $0.5a$, the same radius as a solvent particle used in the DPD stage of the simulation. In all studied simulations, at timescales $0 < t/\tau_D < t/\tau_{D\,delay}$, SASA decreases slightly with time, consistent with the formation of an initial $C_1$ structure (**Fig 3B**). Here, SASA remains high due to the presence of a randomly dispersed $C_2$ population. At $t/\tau_D > t/\tau_{D\,delay}$, however, SASA begins to rapidly decrease, indicating secondary gelation caused by the initiation of $C_2$ interspecies and intraspecies attractions. At $t/\tau_D \gg t/\tau_{D\,delay}$, a complete gel structure is formed, and structural evolution slows down, with SASA slowly reaching a plateau. SASA measurements were normalized by the number of colloidal particles to establish a continuity between the DPD and LD stages of the simulation (refer to Methods for an explanation of the two stages of the simulation), as the number of colloidal particles in these simulations differed by a factor of 27. A slight disconnect is evident in SASA measurements between the end of the DPD stage and the beginning of the LD stage because the utilized algorithm does not account for periodicity, instead treating the contents of each simulation box as a distinct, non-repeating structure. As a result, when the final state of the DPD stage is scaled up across its periodic boundaries, a lower fraction of the particles remains at the edges, thereby artificially decreasing the measured SASA due to interference from their periodic neighbors. Nevertheless, despite this slight discrepancy, SASA is useful for illustrating trends in surface area and coarseness of overall structures over time. Measurements show that in gels with $D_{0\ (1-2)}$ = 10$k_B$T, SASA of the final gel state decreases with increased $t/\tau_{D\ delay}$ (**Fig 3B**), with no discernable differences observed amongst systems with $D_{0\ (1-2)}$ = 5$k_B$T and $D_{0\ (1-2)}$ = 0 (**Fig S1, S2**). Bearing in mind the previously discussed observations regarding pore and cluster size, it becomes evident that in systems with dominant interspecies attraction, the initiation of secondary gelation serves as an inhibitor of coarsening; the longer secondary gelation is delayed, the coarser the gel is able to become. In contrast, systems lacking this dominant interspecies attraction do not display a significant correlation between $t/\tau_{D\ delay}$ and coarseness. In addition to the coarseness measure, and to quantify the extent of mixing/de-mixing between the two colloidal species, we performed a cluster analysis to characterize the dominant structures formed by each colloid, and an average coordination number analysis to illustrate global trends in neighbor selectivity and structural evolution.

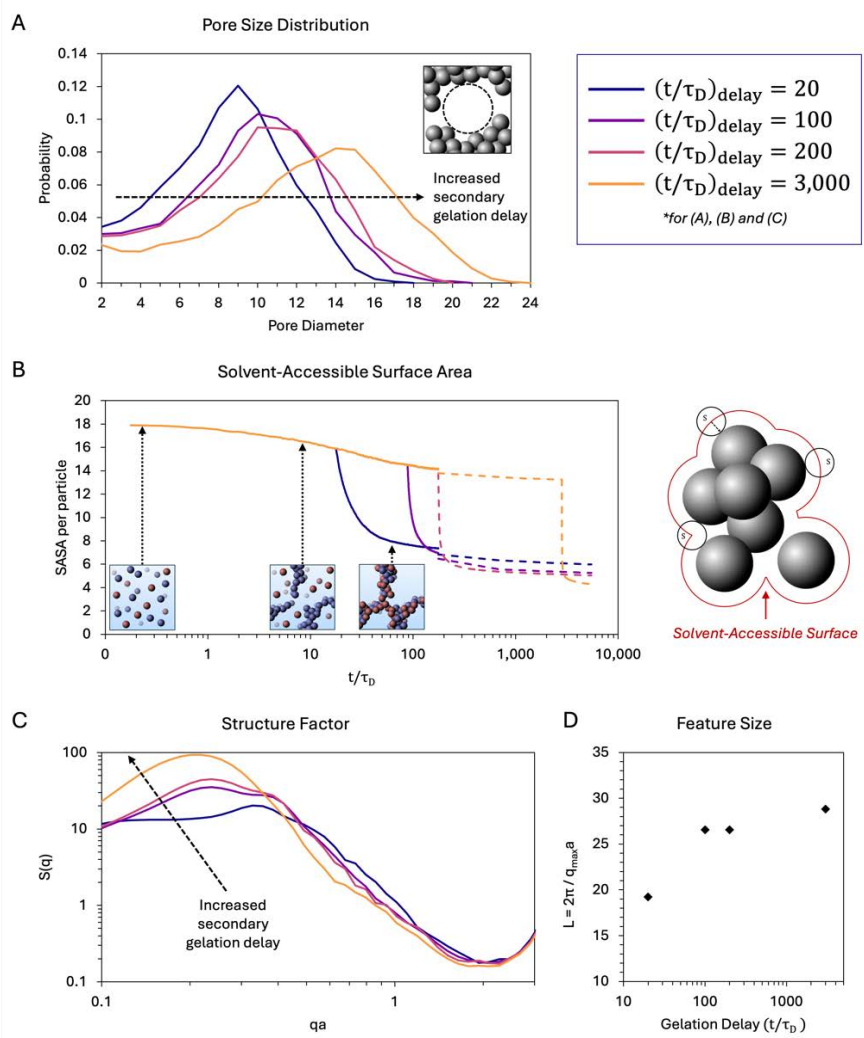

**Figure 3**: Characterization of gels with $D_{0\ (1-2)}$ = 10$k_B$T. (A) Pore size distribution at the end of simulation. (B) Solvent-accessible surface area (normalized per particle) over time. Solid lines represent the DPD stage, dotted lines represent the LD stage. (C) Static structure factor S(q) at the end of the simulation. (D) Characteristic feature size obtained from S(q) measurements.

## 2.2 Species Segregation and Mixing Behavior

In subsequent analyses, we consider a cluster as a group of particles where every particle is within a "bonding distance" of 0.1*a* of at least one other member of said group. In this context, we define the largest connected component (LCC) as the largest cluster present in the system, and $f_{LCC}$ as the fraction of the total colloid population contained within the LCC. In addition to clusters of indiscriminate composition, we also analyze single-species clusters consisting purely of $C_1$ or $C_2$. Similarly, coordination number analysis measures the average number of neighbors within 0.1*a* that each particle has. This metric is applied to three measurements; the mean number of same-species neighbors each $C_1$ particle has ($Z_{1-1}$), the mean number of same-species neighbors each $C_2$ particle has ($Z_{2-2}$), and the mean number of different-species neighbors each particle has ($Z_{1-2}$).

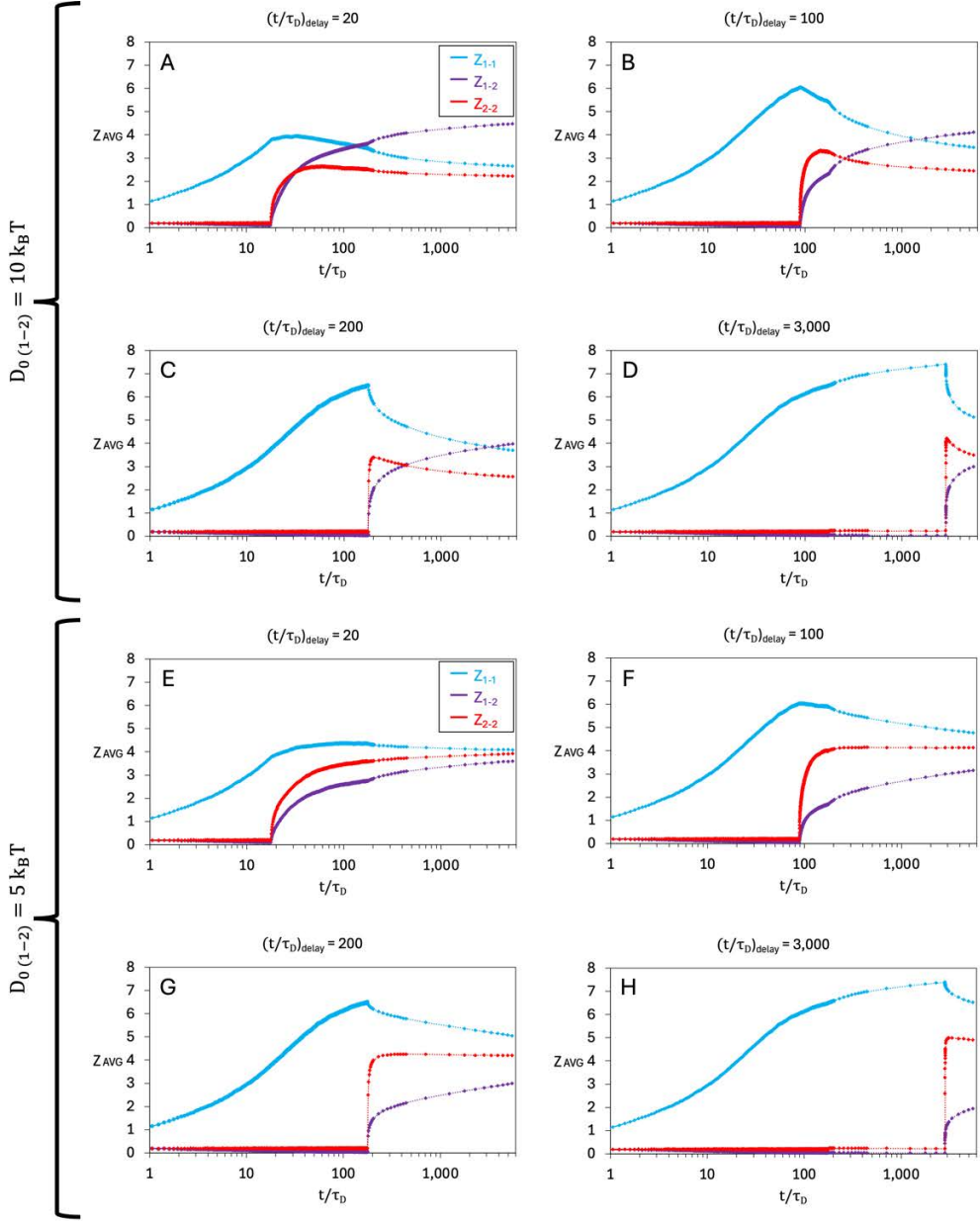

**Figure 4**: Average coordination number analyses for gels with interspecies attraction (A-D) $D_{0\,(1-2)}$ = 10$k_B$T and (E-H) $D_{0\,(1-2)}$ = 5$k_B$T. (A, E) $t/\tau_{D\,delay}$ = 20. (B, F) $t/\tau_{D\,delay}$ = 100. (C, G) $t/\tau_{D\,delay}$ = 200. (D, H) $t/\tau_{D\,delay}$ = 3,000. Blue lines represent

the average number of $C_1$ neighbors a $C_1$ particle has. Red lines represent the average number of $C_2$ neighbors a $C_2$ particle has. Purple lines represent the average number of $C_2$ neighbors a $C_1$ has, and vice-versa.

In all studied gels, $f_{LCC}$ of the overall system was ~100% by the end of the simulations, indicating complete gelation. Additionally, in gels with interspecies attractions $D_{0\ (1-2)}$ = 5$k_B$T and $D_{0\ (1-2)}$ = 0, $f_{LCC}$ for every individual species was ~95% regardless of $t/\tau_{D\ delay}$ at this point. These results are not unexpected. In the case of $D_{0\ (1-2)}$ = 5$k_B$T, interspecies attraction is the same as intraspecies attractions, meaning that particles are not preferentially attracted to one species or another. Because weakly attracted colloidal gels are generally coarse at large timescales(8), most particles possess multiple neighbors and, for any given particle, it is likely that at least one such neighbor belongs to the same species (**Fig 4E-H**). In the case of $D_{0\ (1-2)}$ = 0, the absence of interspecies attraction necessitates the formation of single-species clusters, with the colloid concentration being sufficient to form space-spanning networks of both species.

**Table 1**: Largest Connected Component (LCC) analysis of gels with $D_{0\ (1-2)}$ = 10$k_B$T at the end of simulation.

| $t/\tau_{D\ delay}$ | $LCC_{C1}$ | $LCC_{C2}$ | $f_{C1}$ | $f_{C2}$ |
|---|---|---|---|---|
| 20 | 150,623 | 7,152 | 68.1% | 3.2% |
| 100 | 200,076 | 17,377 | 90.5% | 7.9% |
| 200 | 201,695 | 81,999 | 91.2% | 37.1% |
| 3,000 | 211,125 | 193,150 | 95.5% | 87.4% |

However, single species $f_{LCC}$ becomes a lot more descriptive when comparing gels with $D_{0\ (1-2)}$ = 10$k_B$T. Notably, we observe that as $t/\tau_{D\ delay}$ is increased, $f_{C1}$ and $f_{C2}$ grow as well. Notably, $f_{C2}$ steadily increases from 3% to 87% over the range of tested $t/\tau_{D\ delay}$ values (**Table 1**). Visual observation reveals the emergence of coated structures in systems with $D_{0\ (1-2)}$ = 10$k_B$T at high $t/\tau_{D\ delay}$ values, with $C_2$ particles largely encapsulating a space-spanning internal $C_1$ structure (**Fig 2**). The exception to this occurs at low $t/\tau_{D\ delay}$, where the initial $C_1$ scaffold doesn't have enough time to fully form into a space-spanning network prior to the start of secondary gelation, resulting in a more mixed structure. From this, we conclude that as $t/\tau_{D\ delay}$ is increased, the $C_2$ coating around the initial $C_1$ scaffold becomes more complete. This can be attributed to the prolonged formation of the initial $C_1$ scaffold; the longer secondary gelation is delayed, the more completely formed and/or coarse this $C_1$ scaffold becomes, resulting in a lower surface area. The decreased surface area is then covered more completely by $C_2$ particles. This de-mixing trend is further confirmed by average coordination number analysis (**Fig 4A-D**), which illustrates that as $t/\tau_{D\ delay}$ grows, so too does the average number of intraspecies neighbor pairs, at the expense of interspecies bonding. A similar "de-mixing" phenomenon is observed for gels with $D_{0\ (1-2)}$ = 5$k_B$T. Although LCC analysis is largely uninformative in this case, average coordination number analysis suggests the formation of an initial $C_1$ scaffold that, with increasing $t/\tau_{D\ delay}$, coarsens and prevents increasingly more of its internal $C_1$ constituents from being exposed to $C_2$ neighbors (**Fig 4E-H**). This can be confirmed visually by referring to **Fig 2**, where it is visually evident that as $t/\tau_{D\ delay}$ is increased, the two colloid species become more segregated into distinct $C_1$ core and $C_2$ coating.

## 3 DISCUSSION

Through detailed large-scale simulations, in this work we have demonstrated that in two-component colloidal systems undergoing sequential gelation, by exploiting the extent of intra-particle attraction as well as the length of a secondary gelation delay, a zoology of different structural morphologies can be achieved: from completely de-mixed double networks, to core-shell composite structures, and to fully mixed gels with varying levels of length-scales and structural uniformities. We find that in systems with nonzero interspecies attraction, a short secondary gelation delay leads to the formation of a mixed stricture, whereas an extended secondary gelation delay promotes the formation of a core-shell composite structure, consisting of an internal scaffold made up of the initially gelling species, and an external sheath made of the secondary species. Broadly, the de-mixing of the two colloidal species into distinct "core" and "coating" structures becomes more and more pronounced the longer secondary gelation is delayed, a finding that is consistent with previously published work(25). More unexpectedly, in systems with dominant interspecies interactions, we observe a direct correlation between the length of the secondary gelation delay and the coarseness of the final structure. Here, it appears that the strong interspecies bonds eventually formed between the particles on the exterior of the "core" structures and the interior of the "coating" structures lock in the interior structure shortly after the start of secondary gelation, suppressing its propensity to coarsen over time. Collectively, these findings present compelling implications for the tunability of multicomponent colloidal systems.

It is known that colloidal gels with weak interparticle attractions continue to evolve and coarsen with age(43), even at longer timescales than what has been presented in this study(8). Indeed, although this process becomes quite slow, it is evident that each of the studied gels continue to evolve even at the end of the simulation. That said, there is little indication that the trends in the overall structures observed in this study would not continue to hold even if the simulations were extended further. Globally, this is best evidenced by SASA measurements; in every set of gels of a given interspecies attraction $D_{0\ (1\text{-}2)}$, SASA decreases at approximately the same rate, suggesting a roughly equivalent rate of coarsening. It is, however, plausible that at timescales beyond those studied here, individual species may continue to become more mixed within the gel structure. This is suggested by the gradual decline of same-species $C_1$ neighbors and the rise of mixed-species neighbor pairs (**Fig 4**). Again, this can be attributed to the relative weakness of the interspecies attractions, which make the studied gels prone to continual evolution. Nevertheless, we clearly demonstrate that despite the transient nature of the previously described mixing/de-mixing behavior, it is still evident that the length of the secondary gelation delay continues to have a lasting impact on species segregation, even at times long after the initiation of secondary gelation ($t/\tau_D \gg t/\tau_{D\ delay}$). Our results clearly show that temporal manipulation of attractive interactions between colloids and the extent of attraction between different colloids can result in structures and morphologies that are significantly different from ones formed during a simple gelation process. While the scope of studied systems in this work was deliberately limited to weakly attractive systems, our findings strongly suggest that fabrication of colloidal designer and sacrificial networks can be achieved by leveraging combined controls over particle-level physical interactions and gelation kinetics simultaneously.

## 4 METHODS

### 4.1 Two-Stage Simulation Protocol

Throughout this work, all colloidal systems are computationally studied using large scale simulations, providing a detailed view of particle coordinates and trajectories at all times. Simulations were performed using a modified version of the HOOMD-blue suite of molecular dynamics software(44). The computational studies are divided into two parts: **1)** In the first stage, Dissipative Particle Dynamics (DPD) is used to model the initial gelation of colloidal particles from a randomly dispersed equilibrium configuration. DPD was selected as the method of choice for this stage as it accounts for the hydrodynamics of the system, a quality that is crucial for accurately modelling gelation dynamics(45–48). **2)** In the second stage, Langevin Dynamics (LD) method is used to simulate the aging of resulting particulate assemblies, starting from an initial configuration constructed from scaling up the final state of the first stage. In DPD, fluid particles are modeled explicitly, naturally preserving the full hydrodynamics picture of the system; however, this also makes extremely large-scale simulations of colloidal systems prohibitive. Compared to DPD, LD is capable of simulating vastly more colloidal particles and utilizing longer timesteps, allowing for larger simulations at longer timescales since the fluid phase is implicitly accounted for. We hypothesize that the hydrodynamic effects are most important during the initial stages of gelation, and less influential during the coarsening of the structure. Thus, while LD does not accurately represent the hydrodynamics of individual colloidal particles, we presume that this is less detrimental for systems that have already undergone the initial phase of gelation. Using a sequential combination of these two modelling methods allowed for the advantages of both methods to be exploited: maintaining accurate hydrodynamics at startup, all the while being able to ultimately simulate statistically significant quantities of colloidal particles at long timescales.

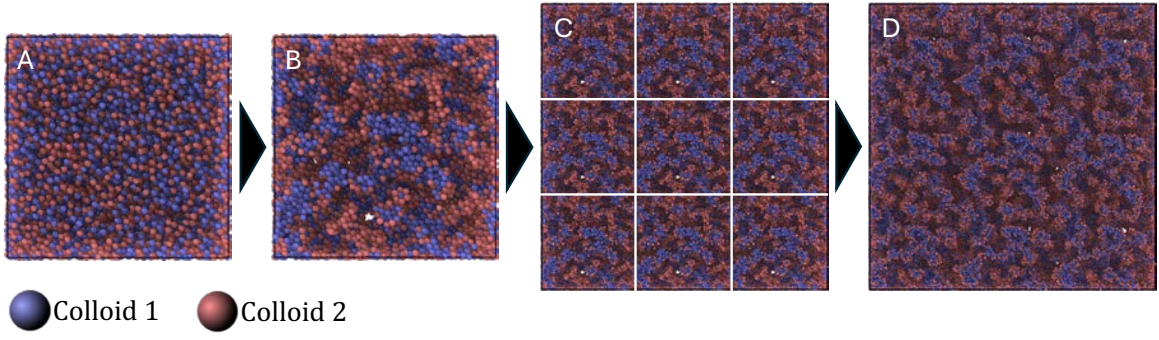

**Figure 5**: Visualization of the simulation process (A) Initial random state (B) Final state obtained from the "gelation" stage, simulated using DPD for 200 $\tau_D$ (C) Periodic box from (B) scaled up by a factor of 3 in each direction, used as the initial state for "post-gelation" stage (D) Final "post-gelation" state, simulated using LD for a further 5,300 $\tau_D$.

       The initial configuration of all simulations consisted of a cubic simulation box of size L=70 with periodic boundary conditions, filled with 20% $_{vol}$ of randomly dispersed colloidal particles split evenly into species $C_1$ and $C_2$ (N=16,378 particles in total) (**Fig 5A**). DPD was used to simulate the initial gelation of this system for 200 $\tau_D$ (**Fig 5B**). At the end of this stage, the final state of the DPD simulation was scaled up by a factor of 3 across its periodic boundaries in every direction (**Fig 5C**). Then, LD was used to simulate a "post-gelation" stage of N=442,206 particles for an additional 5,300 $\tau_D$ (**Fig 5D**). In total, the simulations are run for 5,500 $\tau_D$ between the two stages.

### 4.2 Dissipative Particle Dynamics

       In the utilized modified DPD model, described in detail by Boromand et al.(49), both colloidal and solvent particles are considered. The equation of motion for particles is described as:

$$m_i \frac{\partial \boldsymbol{v_i}}{\partial t} = \sum \boldsymbol{F}_{ij}^C + \boldsymbol{F}_{ij}^D + \boldsymbol{F}_{ij}^R + \boldsymbol{F}_{ij}^{Att} + \boldsymbol{F}_{ij}^H + \boldsymbol{F}_{ij}^{Core}$$

Here, $m_i$ and $\boldsymbol{v}_i$ represent particle mass and velocity, respectively. Solvent-solvent and solvent-colloid interactions are described by the sum of the first three terms on the right side: conservative force $\boldsymbol{F}_{ij}^C$, representing a soft repulsive potential based on solvent compressibility(50) ($\boldsymbol{F}_{ij}^C = a_{ij}\omega_{ij}^C \boldsymbol{e}_{ij}$, where $a_{ij}$ is the maximum repulsion coefficient, $\omega_{ij}^C$ is a distance-dependent weight function, and $\boldsymbol{e}_{ij}$ is the unit normal vector); dissipative force $\boldsymbol{F}_{ij}^D$, representing viscous effects ($\boldsymbol{F}_{ij}^D = -\gamma_{ij}\omega_{ij}^D(\boldsymbol{v}_{ij} \cdot \boldsymbol{e}_{ij})\boldsymbol{e}_{ij}$, where $\gamma_{ij}$ is the drag coefficient, $\omega_{ij}^D$ is a distance-dependent weight function, $\boldsymbol{v}_{ij}$ is the relative velocity between particles); and random force $\boldsymbol{F}_{ij}^R$, representing Brownian thermal fluctuations ($\boldsymbol{F}_{ij}^R = (\sigma_{ij}\omega_{ij}^R \Theta_{ij}/\sqrt{\Delta t})\boldsymbol{e}_{ij}$, where $\sigma_{ij}$ is the random force strength, $\omega_{ij}^R$ is a distance-dependent weight function, $\Theta_{ij}$ is a Gaussian random parameter with a mean of 0, and $\Delta t$ is timestep size). In the aforementioned forces, weight functions are dependent on relative position $r_{ij}$ and are defined as $\omega_{ij}^C = \omega_{ij}^R = (\omega_{ij}^R)^{0.5} = 1 - (r_{ij}/r_c)$ when the particles are within a cutoff distance $r_{ij} < r_c$ away from each other, and as $\omega_{ij}^C = \omega_{ij}^R = (\omega_{ij}^R)^{0.5} = 0$ otherwise. Colloid-colloid interactions are described by the sum of the remaining forces: attractive force $\boldsymbol{F}_{ij}^{Att}$, hydrodynamic force $\boldsymbol{F}_{ij}^H$ and core potential $\boldsymbol{F}_{ij}^{Core}$. Here, $\boldsymbol{F}_{ij}^{Att}$ represents short-range interaction, modelled using Morse potential (**Fig 1D**). In cases where colloidal particle pairs do not experience such short-range attractions to each other (that is, when $D_0 = 0$). $\boldsymbol{F}_{ij}^{Core}$ represents a semi-hard core potential interaction used to prevent non-physical overlap of non-interacting colloidal particle pairs upon contact (modelled as the "repulsive" half of a Morse potential where $h_{ij} < 0$, that is only invoked when two non-interacting colloidal particles overlap, as this force would be redundant with $\boldsymbol{F}_{ij}^{Att}$ in the case of interacting colloidal particles overlapping). The numerical definition of $\boldsymbol{F}_{ij}^{Core}$ is the only deviation from the model as described by Boromand et al.(49). Lastly, $\boldsymbol{F}_{ij}^H$ represents short-range lubrication effects in cases where the distance between two colloidal particles is too small for a solvent particle to fit between them ($F_{ij}^H = -\mu_{ij}^H(v_{ij} \cdot e_{ij})e_{ij}$, where pair drag term $\mu_{ij}^H = 3\pi\eta a_1 a_2/2h_{ij}$, and $a_1$ and $a_2$ represent colloidal particle radii. Below a sufficiently small cutoff interparticle distance $h_{ij} < \delta$, the value of $h_{ij}$ is substituted with $\delta$ in order to prevent divergence of the drag term to infinity). Integration over time is performed with the Velocity-Verlet method (11, 50).

**4.3 Langevin Dynamics**

The LD model utilized in the study was included as-is in the HOOMD-blue suite(44) and was not subject to any modification. In this model, only colloidal particles are considered, and the solvent is modeled implicitly. The motion of the simulated particles is described by the following equation:

$$m_i \frac{\partial v_i}{\partial t} = F_C - \gamma \cdot v_i + F_R$$

Here, $F_C$ is the total force exerted on a particle from all colloid-colloid interaction potentials and constraint forces (here modelled using Morse potential in a manner analogous to $F_{ij}^{Att}$ and $F_{ij}^{Core}$ in our DPD method); $\gamma$ is the drag coefficient; and $F_R$ is a random force with a magnitude consistent with $\gamma$ and temperature $T$, meant to represent Brownian thermal fluctuations. Similarly to DPD, integration over time was performed via the Velocity-Verlet method(51, 52).

**ACKNOWLEDGEMENTS**

A.K. and S.J. acknowledge support by the Office of Naval Research through Award N000142312772. Computational resources were provided by the Massachusetts Green High Performance Computing Center in Holyoke, MA.

**AUTHOR CONTRIBUTIONS**

Conceptualization, S.J.; Methodology, A.K. and S.J.; Investigation, A.K. and S.J.; Writing–original draft, A.K. and S.J.; Writing–review & editing, A.K. and S.J.; Funding acquisition, S.J.; Resources, S.J.; Supervision, S.J.

**COMPETING INTERESTS**

The authors declare no competing interests.

**MATERIALS & CORRESPONDENCE**

Requests for materials and correspondence should be addressed to Safa Jamali (s.jamali@northeastern.edu).

**REFERENCES**

1. Y. Cao, R. Mezzenga, Design principles of food gels. *Nat. Food* **1**, 106–118 (2020).

2. R. M. Martinez, C. Rosado, M. V. R. Velasco, S. C. S. Lannes, A. R. Baby, Main features and applications of organogels in cosmetics. *Int. J. Cosmet. Sci.* **41**, 109–117 (2019).

3. N. Peppas, Hydrogels in pharmaceutical formulations. *Eur. J. Pharm. Biopharm.* **50**, 27–46 (2000).

4. J.-L. Barrat, *et al.*, Soft matter roadmap. *J. Phys. Mater.* **7**, 012501 (2024).

5. P. J. Lu, *et al.*, Gelation of particles with short-range attraction. *Nature* **453**, 499–503 (2008).

6.  M. E. Helgeson, S. E. Moran, H. Z. An, P. S. Doyle, Mesoporous organohydrogels from thermogelling photocrosslinkable nanoemulsions. *Nat. Mater.* **11**, 344–352 (2012).

7.  M. Bantawa, *et al.*, The hidden hierarchical nature of soft particulate gels. *Nat. Phys.* **19**, 1178–1184 (2023).

8.  R. N. Zia, B. J. Landrum, W. B. Russel, A micro-mechanical study of coarsening and rheology of colloidal gels: Cage building, cage hopping, and Smoluchowski's ratchet. *J. Rheol.* **58**, 1121–1157 (2014).

9.  M. Nabizadeh, S. Jamali, Life and death of colloidal bonds control the rate-dependent rheology of gels. *Nat. Commun.* **12**, 4274 (2021).

10. D. Mangal, S. Jamali, Role of interaction range on the microstructure and dynamics of attractive colloidal systems. *Soft Matter* **20**, 4466–4473 (2024).

11. S. Jamali, R. C. Armstrong, G. H. McKinley, Time-rate-transformation framework for targeted assembly of short-range attractive colloidal suspensions. *Mater. Today Adv.* **5**, 100026 (2020).

12. F. J. Müller, L. Isa, J. Vermant, Toughening colloidal gels using rough building blocks. *Nat. Commun.* **14**, 5309 (2023).

13. N. Dagès, *et al.*, Interpenetration of fractal clusters drives elasticity in colloidal gels formed upon flow cessation. *Soft Matter* **18**, 6645–6659 (2022).

14. J. P. Gong, Y. Katsuyama, T. Kurokawa, Y. Osada, Double-Network Hydrogels with Extremely High Mechanical Strength. *Adv. Mater.* **15**, 1155–1158 (2003).

15. Y. Yin, Q. Gu, X. Liu, F. Liu, D. J. McClements, Double network hydrogels: Design, fabrication, and application in biomedicines and foods. *Adv. Colloid Interface Sci.* **320**, 102999 (2023).

16. C. Ferreiro-Córdova, E. Del Gado, G. Foffi, M. Bouzid, Multi-component colloidal gels: interplay between structure and mechanical properties. *Soft Matter* **16**, 4414–4421 (2020).

17. X. Guo, *et al.*, Robust zwitterionic hydrogels enabled by consolidated supramolecular networks and spatially hierarchical structures. *Nat. Commun.* **16**, 9454 (2025).

18. J.-Y. Sun, *et al.*, Highly stretchable and tough hydrogels. *Nature* **489**, 133–136 (2012).

19. X. Liu, *et al.*, Topoarchitected polymer networks expand the space of material properties. *Nat. Commun.* **13**, 1622 (2022).

20. Z. J. Wang, *et al.*, Rapid self-strengthening in double-network hydrogels triggered by bond scission. *Nat. Mater.* **24**, 607–614 (2025).

21. F. Varrato, *et al.*, Arrested demixing opens route to bigels. *Proc. Natl. Acad. Sci.* **109**, 19155–19160 (2012).

22. M. L. Mugnai, R. Tchuenkam Batoum, E. Del Gado, Interspecies interactions in dual, fibrous gels enable control of gel structure and rheology. *Proc. Natl. Acad. Sci.* **122**, e2423293122 (2025).

23. E. R. Draper, E. G. B. Eden, T. O. McDonald, D. J. Adams, Spatially resolved multicomponent gels. *Nat. Chem.* **7**, 848–852 (2015).

24. J. N. Immink, J. J. E. Maris, J. J. Crassous, J. Stenhammar, P. Schurtenberger, Reversible Formation of Thermoresponsive Binary Particle Gels with Tunable Structural and Mechanical Properties. *ACS Nano* **13**, 3292–3300 (2019).

25. J. Appel, *et al.*, Temperature Controlled Sequential Gelation in Composite Microgel Suspensions. *Part. Part. Syst. Charact.* **32**, 764–770 (2015).

26. L. Di Michele, *et al.*, Multistep kinetic self-assembly of DNA-coated colloids. *Nat. Commun.* **4**, 2007 (2013).

27. A. C. Lima, C. R. Correia, M. B. Oliveira, J. F. Mano, Sequential ionic and thermogelation of chitosan spherical hydrogels prepared using superhydrophobic surfaces to immobilize cells and drugs. *J. Bioact. Compat. Polym.* **29**, 50–65 (2014).

28. A. Abu-Hakmeh, *et al.*, Sequential gelation of tyramine-substituted hyaluronic acid hydrogels enhances mechanical integrity and cell viability. *Med. Biol. Eng. Comput.* **54**, 1893–1902 (2016).

29. R. Luo, Y. Cao, P. Shi, C. Chen, Near-Infrared Light Responsive Multi-Compartmental Hydrogel Particles Synthesized Through Droplets Assembly Induced by Superhydrophobic Surface. *Small* **10**, 4886–4894 (2014).

30. Q. Zhu, *et al.*, Design and structural characterization of edible double network gels based on wheat bran arabinoxylan and pea protein isolate. *Int. J. Biol. Macromol.* **213**, 824–833 (2022).

31. J. Guo, *et al.*, Fabrication of edible gellan gum/soy protein ionic-covalent entanglement gels with diverse mechanical and oral processing properties. *Food Res. Int.* **62**, 917–925 (2014).

32. B. Tavsanli, V. Can, O. Okay, Mechanically strong triple network hydrogels based on hyaluronan and poly(N,N-dimethylacrylamide). *Soft Matter* **11**, 8517–8524 (2015).

33. M. Grzelczak, J. Vermant, E. M. Furst, L. M. Liz-Marzán, Directed Self-Assembly of Nanoparticles. *ACS Nano* **4**, 3591–3605 (2010).

34. D. Guo, Y. Song, Recent Advances in Multicomponent Particle Assembly. *Chem. – Eur. J.* **24**, 16196–16208 (2018).

35. E. Zaccarelli, *et al.*, Crystallization of Hard-Sphere Glasses. *Phys. Rev. Lett.* **103**, 135704 (2009).

36. P. N. Pusey, *et al.*, Hard spheres: crystallization and glass formation. *Philos. Trans. R. Soc. Math. Phys. Eng. Sci.* **367**, 4993–5011 (2009).

37. S. I. Henderson, T. C. Mortensen, S. M. Underwood, W. Van Megen, Effect of particle size distribution on crystallisation and the glass transition of hard sphere colloids. *Phys. Stat. Mech. Its Appl.* **233**, 102–116 (1996).

38. A. Filipponi, The radial distribution function probed by X-ray absorption spectroscopy. *J. Phys. Condens. Matter* **6**, 8415–8427 (1994).

39. V. Ramasubramani, *et al.*, freud: A software suite for high throughput analysis of particle simulation data. *Comput. Phys. Commun.* **254**, 107275 (2020).

40. C. D. Muzny, B. D. Butler, H. J. M. Hanley, Evolution of the Structure Factor in Gelling Dense Colloidal Silica. *MRS Proc.* **407**, 87 (1995).

41. L. D. Gelb, K. E. Gubbins, Pore Size Distributions in Porous Glasses: A Computer Simulation Study. *Langmuir* **15**, 305–308 (1999).

42. W. Humphrey, A. Dalke, K. Schulten, VMD: Visual molecular dynamics. *J. Mol. Graph.* **14**, 33–38 (1996).

43. R. Buscall, *et al.*, Towards rationalising collapse times for the delayed sedimentation of weakly-aggregated colloidal gels. *Soft Matter* **5**, 1345 (2009).

44. J. A. Anderson, J. Glaser, S. C. Glotzer, HOOMD-blue: A Python package for high-performance molecular dynamics and hard particle Monte Carlo simulations. *Comput. Mater. Sci.* **173**, 109363 (2020).

45. Z. Varga, J. Swan, Hydrodynamic interactions enhance gelation in dispersions of colloids with short-ranged attraction and long-ranged repulsion. *Soft Matter* **12**, 7670–7681 (2016).

46. A. Furukawa, H. Tanaka, Key Role of Hydrodynamic Interactions in Colloidal Gelation. *Phys. Rev. Lett.* **104**, 245702 (2010).

47. J. De Graaf, W. C. K. Poon, M. J. Haughey, M. Hermes, Hydrodynamics strongly affect the dynamics of colloidal gelation but not gel structure. *Soft Matter* **15**, 10–16 (2019).

48. S. Jamali, J. F. Brady, Alternative Frictional Model for Discontinuous Shear Thickening of Dense Suspensions: Hydrodynamics. *Phys. Rev. Lett.* **123**, 138002 (2019).

49. A. Boromand, S. Jamali, J. M. Maia, Structural fingerprints of yielding mechanisms in attractive colloidal gels. *Soft Matter* **13**, 458–473 (2017).

50. R. D. Groot, P. B. Warren, Dissipative particle dynamics: Bridging the gap between atomistic and mesoscopic simulation. *J. Chem. Phys.* **107**, 4423–4435 (1997).

51. M. P. Allen, D. J. Tildesley, *Computer simulation of liquids*, 2nd ed (Oxford university press, 2017).

52. D. J. Evans, G. Morriss, *Statistical Mechanics of Nonequilibrium Liquids*, 2nd Ed. (Cambridge University Press, 2008).